\documentclass[5p, a4paper]{elsarticle}

\usepackage{amsmath}
\usepackage{amsfonts}
\usepackage{amssymb}
\usepackage{mathtools}
\usepackage{dsfont}
\usepackage{marginnote}
\usepackage{pgfplots}
\usepackage{pgfplotstable}
\usepgfplotslibrary{patchplots}
\usepgfplotslibrary{groupplots}
\usetikzlibrary{arrows, automata, positioning}

\pgfplotsset{compat=1.8,
  colormap={whitered}{color(0cm)=(white); rgb255(2cm)=(153, 153, 255)}
}

\newtheorem{theorem}{Theorem}
\newtheorem{assumption}[theorem]{Assumption}
\newtheorem{definition}[theorem]{Definition}
\newtheorem{example}[theorem]{Example}

\newtheorem{remark}[theorem]{Remark}

\newtheorem{proposition}[theorem]{Proposition}

\numberwithin{theorem}{section}

\newenvironment{proof}{{\em Proof.}}{\hfill \hspace*{1pt}
\hfill $\blacksquare$}

\newcommand\RE{\mathbb{R}}

\newcommand\dom{\textrm{Dom}}

\begin{document}

\begin{frontmatter}
 \title{Differential dissipativity analysis of  
 reaction-diffusion systems\tnoteref{erc}}

 \author{Felix A. Miranda-Villatoro} \ead{fam48@cam.ac.uk}
 \author{Rodolphe Sepulchre} \ead{r.sepulchre@eng.cam.ac.uk}
 \address{Engineering Department, University of Cambridge, UK}

 \tnotetext[erc]{The research leading to these
   results has received funding from the European Research Council
   under the Advanced ERC Grant Agreement Switchlet n.670645.}

\begin{abstract}
This note shows how classical tools from linear control theory can be leveraged
 to provide a global analysis of nonlinear reaction-diffusion models. 
The approach is differential in nature. It proceeds from classical tools
of contraction analysis and recent extensions to differential dissipativity.

\end{abstract}

\end{frontmatter}

\section{Introduction}

Reaction-diffusion equations are broadly used for modeling the 
spatio-temporal evolution of processes appearing in many fields 
of science such as propagation of electrical activity on cells in
cellular biology \cite{keener2009}; reactions between substances on active media in 
chemistry \cite{kuramoto1984}; transport phenomena in semiconductor devices in
electronics \cite{scholl2001}; and combustion processes and heat propagation in 
physics \cite{temam1997}, to name a few. 
They have attracted recent interest in control, most notably  \cite{aminzare2016} and
\cite{arcak2011}, because the close link between reaction-diffusion systems
and synchronization models under diffusive coupling: 
the linear diffusion term in reaction-diffusion
partial differential equations is the continuum limit of the diffusive (or incrementally
passive) interconnection network of agents sharing the same
reaction dynamics. In that sense, the results in \cite{aminzare2016} and
\cite{arcak2011} are infinite dimensional generalizations of classical 
finite dimensional results pertaining to synchronization 
\cite{proskurnikov2015, slotine2005, stan2007, vandeWouw2006}.

Our contribution in the present note is to further emphasize the potential of
classical tools from linear control theory in the analysis and design of reaction-diffusion
systems. Our observation is twofold. First, we model reaction diffusion systems as
the interconnection of a linear spatially and time-invariant (LTSI) model with
a static nonlinearity. This natural decomposition calls for a dissipativity
analysis of the interconnection, with complementary input-output dissipation 
inequalities imposed on the LTSI model and on the static nonlinearity, respectively.
Second, we study this interconnection {\it differentially} along arbitrary solutions, 
thereby studying a nonlinear model through a family on linearized systems.

The proposed approach is largely inspired from \cite{aminzare2016} and
\cite{arcak2011}, which analyze spatial homogeneity via contraction theory.
The purely differential approach in the present paper is thought to
offer further potential especially in situations where the attractor is difficult
to characterize explicitely.  In this note, we illustrate 
the benefits of a differential approach in two distinct ways:  (i) we use the classical
KYP lemma to complement existing state-space analysis results with a frequency-domain analysis;
and (ii) we use recent results of differential dissipativity theory \cite{forni2019} to characterize the
attractor of two classical reaction-diffusion models: Nagumo model of bistability 
\cite{nagumo1965}, and Fitzugh-Nagumo model of oscillation \cite{fitzhugh1961}.

\subsection*{Some notation}
Let $L_{n}^{2}(\Omega)$ denote the Hilbert space of square integrable 
functions mapping $\Omega \subset \RE$ to $\RE^{n}$ with the conventional inner
product $\langle x, y \rangle_{L_{n}^{2}(\Omega)}
= \int_{\Omega} x(\theta)^{\top} y(\theta) d \theta$
and norm denoted by $\Vert \cdot \Vert_{L_{n}^{2}(\Omega)}$. 
When clear from the context, we will drop the subindex. For 
vectors $\xi, \psi$ in $\RE^{n}$, 
the inner product is denoted as $\xi^{\top} \psi$ and the
associated norm as $\vert \cdot \vert$. 
The set $\mathbb{C}_{+} := \{a + j b \in \mathbb{C} | a \geq 0 \}$ 
denotes the set of complex numbers with non-negative real part, whereas $\RE_{+}$
denotes the set of non-negative real numbers. A symmetric,
positive (semi-) definite matrix $\Pi$ is denoted as ($\Pi \succeq 0$) $\Pi \succ 0$,
whereas,
$I_{n}$ represents the identity matrix of dimension $n$.

\section{Reaction-diffusion systems}
\label{sec:rd}

The family of distributed systems under consideration has the form
\begin{displaymath}
	\frac{\partial x}{\partial t}(\theta,t) = D \Delta x(\theta,t) + A x(\theta,t)
	- B \varphi(C x(\theta,t))
\end{displaymath}
where $x(\theta,t) \in \RE^{n}$ denotes the state of the system at position 
$\theta \in \Omega \subset \RE$ and time $t \geq 0$. The nonlinear function
$\varphi: \RE^{m} \to \RE^{m}$ represents a static nonlinearity and 
its properties are stated below. 
Spatial diffusion is modeled via the matrix 
$D \in \RE^{n \times n}$ which is symmetric and positive definite, and the 
Laplace operator $\Delta: \dom (\Delta) \subset L_{n}^{2}(\Omega) 
\to L_{n}^{2}(\Omega)$ with domain specified below,
whereas the matrices $A, B$ and $C$ are constant and of the appropriate dimensions.
 
Here, we regard reaction-diffusion systems as the feedback interconnection 
of a linear  system and a static nonlinearity:
\begin{subequations}
 \begin{align}
\Sigma: & \begin{cases}
    \frac{\partial x}{\partial t}(\theta,t) = 
    D \Delta x(\theta,t) + A x(\theta,t) + B u(\theta,t)
    \\
    y(\theta,t) = C x(\theta,t)
  \end{cases}
  \label{eq:pde:sti}
  \\
  & u(\theta,t) = -\varphi(y(\theta,t))
  \label{eq:pde:feedback}
\end{align}
  \label{eq:pde:lure}
\end{subequations}
where $u(\theta,t) \in \RE^{m}$, $y(\theta,t) \in \RE^{m}$ 
are the distributed input and output, respectively. 
For simplicity  we consider the spatial domain $\Omega \subset \RE$ 
as the boundary of the unit circle $S^{1}$.
Thus, $\theta \in [0, 2 \pi]$ and we have the following periodic boundary conditions
\begin{subequations}
 \begin{align}
   \label{eq:pde:bc:1}
  x (0, t) & = x (2 \pi, t)
  \\
  \frac{\partial x}{\partial \theta} (0, t) & = 
  \frac{\partial x}{\partial \theta} (2 \pi, t) 
  \label{eq:pde:bc:2}
\end{align}
 \label{eq:pde:bc}
\end{subequations}

Such models on the circle find application in computational biology for modeling
directional sensing in living cells and systems 
with symmetries see, e.g., \cite{crawford1991, otsuji2007, subramanian2004}.
Additionally, a compact domain simplifies some of the technical assumptions. For instance, 
it guarantees a discrete frequency spectrum for the Laplace operator with domain
 
\begin{equation}
  \label{eq:laplace:domain}
  \dom (\Delta) = \{ x(\cdot, t) \in H^{2}(\Omega; \RE^{n}) | \; 
  \eqref{eq:pde:bc} \text{ holds} \}.
\end{equation}
 
Here $H^{2}(\Omega;\RE^{n}) = H^{2}(\Omega) \times \dots \times H^{2}(\Omega)$ 
denotes the Sobolev space of functions in $L_{n}^{2}(\Omega)$ 
such that the $i$-th component $x_{i}(\cdot, t)$ is differentiable 
(in the generalized sense) with derivatives in $L_{2}(\Omega)$.

The static nonlinearity
$\varphi:\RE^{m} \to \RE^{m}$ is assumed to be 
continuously differentiable (i.e., $\varphi \in C^{1}(\RE^{n}))$ and
satisfies the following standing assumption.
\begin{assumption}
  \label{assump:dissipativeness}
\hfill
  \begin{enumerate}
  \item There exists $0 < M < \infty$ such that $\eta^{\top} \varphi(\eta) < 0$ 
	  for all $\eta \in \RE^{m}$ with $\Vert \eta \Vert \geq M$.
	  
  \item The function $\varphi: \RE^{m} \to \RE^{m}$ satisfies
the differential dissipation inequality
    \begin{equation}
      \begin{bmatrix}
	I_{m} \\ -J_{\varphi}(\eta) 
      \end{bmatrix}^{\top}
      \begin{bmatrix}
	Q & L
	\\
	L^{\top} & R
      \end{bmatrix}
      \begin{bmatrix}
	I_{m} \\ -J_{\varphi}(\eta) 
      \end{bmatrix} \preceq 0
      \label{eq:dissipation:nonlinearity}
    \end{equation}
for all $\eta \in \RE^{m}$, where $J_{\varphi}(\eta) \in \RE^{m \times m}$
denotes the Jacobian matrix of $\varphi$ at $\eta$, the matrices 
$Q, L, R \in \RE^{m \times m}$ are constant and $R = R^{\top} \succ 0$.
  \end{enumerate}
\end{assumption}

The dissipation inequality \eqref{eq:dissipation:nonlinearity} is a classical 
 differential {\it sector condition}. In the scalar case ($m=1$), it reduces to 
  \begin{equation}
\label{eq:diff:sector}
  (J_{\varphi}(\eta) - K_{1})^{\top}(J_{\varphi}(\eta) - K_{2}) \preceq 0
\end{equation}
with $Q = \frac{1}{2} \left( K_{1}^{\top} K_{2} + K_{2}^{\top} K_{1} \right)$, 
$L = \frac{1}{2} (K_{1} + K_{2})^{\top}$ and $R = I_{m}$. Condition \eqref{eq:diff:sector} 
then expresses that the
slope of $\varphi$ at any point lies in the interval $[K_{1}, K_{2}]$,
whenever $K_{1} < K_{2}$. See Figure \ref{fig:n:shape} for an illustration.

The reader will note that 
model  \eqref{eq:pde:lure} reduces to
\begin{equation}
  \frac{\partial x}{\partial t} (\theta,t) = D \Delta x(\theta,t) - 
  \varphi(x(\theta,t))
  \label{eq:pde:rd}
\end{equation}
in the special case defined by $m = n$, $A = 0$, and $B = C = I_{n}$. 
This latter form  is the classical form of a reaction-diffusion system in the literature
\cite{robinson2001}.

\begin{remark}
Assumption \ref{assump:dissipativeness}  ensures that 
the system \eqref{eq:pde:lure}-\eqref{eq:pde:bc} admits a 
unique (classical) solution for any initial condition 
$x(\theta, 0) = x_{0}(\theta)$ which is defined in the whole time interval
$t \in [0, +\infty)$, given as
\begin{displaymath}
  x(\theta,t) = T(t) x_{0}(\theta) + \int_{0}^{t} T(t - \tau) F(x(\theta, \tau)) d\tau
\end{displaymath}
where $T(t): L_{n}^{2}(\Omega) \to L_{n}^{2}(\Omega)$ is the $C_{0}$-semigroup 
generated by the operator $D \Delta$ and $F(x(\theta, t)) = A x(\theta, t)
+ B \varphi(C x(\theta, t))$. 
See e.g., \cite[Theorems 1.4 and 1.5, Chapter 6]{pazy1983}. 
  \label{remark:existence}
\end{remark}

\section{Differential analysis of reaction diffusion systems}
\label{sec:diff}

Differential analysis consists in analyzing the properties
of infinitesimal variations $\delta x(\theta,t)$ around an arbitrary
solution $x(\theta,t)$ of \eqref{eq:pde:lure}-\eqref{eq:pde:bc}
as is made in \cite{crouch1987, sontag2006} for the case of finite-dimensional systems.

Namely, let $\phi(\theta,t,x_{0})$ denote the solution
of the reaction-diffusion system
\eqref{eq:pde:lure}-\eqref{eq:pde:bc} at position $\theta$ and time $t$
with initial condition $x(\theta,0) = x_{0}(\theta)$.
Let $x^{1}(\theta,0)$ and $x^{2}(\theta,0)$ be two given initial conditions and let 
$\gamma: S^{1} \times [0, 1] \to \RE^{n}$ be a smooth curve such that
$\gamma(\cdot, 0) = x^{1}(\cdot,0)$, $\gamma(\cdot,1) = x^{2}(\cdot,0)$ and 
$\gamma(0, s) = \gamma(2 \pi, s)$ for all $s \in [0, 1]$. In addition, let 
$\psi(\theta,t,s) = \phi(\theta,t,\gamma_{s})$, where $\gamma_{s}(\cdot) = 
\gamma(\cdot, s)$, i.e., $\psi(\theta,t,s)$ is the solution of 
\eqref{eq:pde:lure}-\eqref{eq:pde:bc} at position $\theta$ and time $t$
with initial condition $\psi(\theta, 0, s) = \gamma_{s}(\theta) = \gamma(\theta,s)$, 
$s \in [0, 1]$. It follows that
\begin{multline*}
	\frac{\partial}{\partial t}
	\left( \frac{\partial \psi}{\partial s} (\theta,t,s) \right) = 
	\frac{\partial}{\partial s} \left( \frac{\partial \psi}{\partial t} (\theta,t,s)
	\right)
	\\
	= 
	\frac{\partial}{\partial s} \left(D \Delta \psi(\theta, t, s) + A \psi(\theta,t,s)
	\right.
	\\
	\left. - B \varphi(C \psi(\theta,t,s)) \right)
	\\
	= 
	D \Delta \frac{\partial \psi}{\partial s}(\theta, t, s) + A 
	\frac{\partial \psi}{\partial s}(\theta,t,s) 
	\\
	- B J_{\varphi}(C \psi(\theta,t,s)) C 
	\frac{\partial \psi}{\partial s}(\theta,t,s)
\end{multline*}
Defining 
$\delta x(\theta, t,s):= \frac{\partial \psi}{\partial s}(\theta,t,s)$, leads to
the variational equation
\begin{multline*}
	\frac{\partial \delta x}{\partial t}(\theta,t,s) = D \Delta \delta x(\theta,t,s) 
	+ A \delta x(\theta,t,s) 
	\\
	- B J_{\varphi}(C \psi(\theta,t,s)) C \delta x(\theta,t,s)
\end{multline*}
Equivalently, in Lur'e form

\begin{subequations}
  \begin{align}
    \label{eq:pde:diff:sti}
    \delta \Sigma & :
  \begin{cases}
    \frac{\partial \delta x}{\partial t}(\theta, t) = D \Delta \delta x(\theta, t)
    + A \delta x(\theta, t) + B \delta u(\theta, t)
    \\
    \delta y(\theta, t) = C \delta x(\theta, t)
  \end{cases}
  \\
  & \; \delta u(\theta, t) = - J_{\varphi}(y(\theta, t)) \delta y(\theta, t)
  \label{eq:pde:diff:feedback}
  \end{align}
    \label{eq:pde:diff}
\end{subequations}
with boundary conditions
\begin{subequations}
  \begin{align}
   \delta x(0, t) & = \delta x(2 \pi, t)
   \label{eq:pde:diff:bc:1}
  \\
  \frac{\partial \delta x}{\partial \theta} (0 ,t) & = 
  \frac{\partial \delta x}{\partial \theta} (2 \pi, t)
  \label{eq:pde:diff:bc:2}
  \end{align}
    \label{eq:pde:diff:bc}
\end{subequations}
The variational system is linear. It is the interconnection of the same LTSI
model \eqref{eq:pde:lure} with a time-varying output feedback gain evaluated along
an arbitrary solution $x(\theta, t)$.  In the following subsections we  focus on the analysis of the differential model 
\eqref{eq:pde:diff}-\eqref{eq:pde:diff:bc}. We analyze
spatial and temporal variations separately.

\subsection{Differential inhomogeneous dynamics}

The spatial infinitesimal variation of the solution $x(\theta, t)$ at time $t$ is 
\begin{displaymath}
  \lim_{\nu \to 0} 
  \frac{x(\theta + \nu, t) - x(\theta, t)}{\nu} = 
   \frac{\partial x}{\partial \theta}(\theta, t)
\end{displaymath}
Note that $\frac{\partial x}{\partial \theta}$ is an infinitesimal variation for the family of curves
$\gamma(\theta,s) = \gamma(\theta + s)$. Thus, $\frac{\partial x}{\partial \theta}$ satisfies 
\eqref{eq:pde:diff}-\eqref{eq:pde:diff:bc}.
In addition, it follows from \eqref{eq:pde:bc:1} and the fundamental theorem of
calculus that $\frac{\partial x}{\partial \theta}$ satisfies the integral constraint 
  \begin{equation}
   \label{eq:pde:diff:orthogonal}
   \frac{1}{2 \pi} \int_{0}^{2 \pi} \frac{\partial x}{\partial \theta}(\theta,t) \; d \theta = x(2 \pi, t) - x(0,t)
   = 0,
\end{equation}
which means that $\frac{\partial x}{\partial \theta}$ is a zero mean solution of 
\eqref{eq:pde:diff}-\eqref{eq:pde:diff:bc}. 
Let us consider the bounded linear operator 
$T: L_{n}^{2}(\Omega) \to L_{n}^{2}(\Omega)$
mapping
\begin{equation}
  \delta x \mapsto \int_{\Omega} 
\delta x (\theta, t) d \theta
=: \overline{\delta x}
  \label{eq:projection}
\end{equation}
where $\overline{\delta x}: \RE_{+} \to \RE^{n}$ is a function dependent on time but no longer
dependent on the spatial variable. 
More generally, any 
variation $\delta x$ admits the decomposition
\begin{displaymath}
	\delta x = \overline{\delta x} + \delta \xi
\end{displaymath}
with $\delta \xi \in \mathcal{N}(T)$ and $\overline{\delta x} \in \mathcal{R}(T)$, where
$\mathcal{N}(T)$ and $\mathcal{R}(T)$ denote the null and range space of $T$, respectively.
It follows from \eqref{eq:pde:diff:orthogonal} and the uniqueness of the splitting
that $\delta \xi = \frac{\partial x}{\partial \theta} \in \mathcal{N}(T)$. 
Motivated by the discussion above, we label the dynamics associated to 
$\overline{\delta x} = T \delta x$ as the spatially homogeneous dynamics, 
whereas we refer to the dynamics associated to $\delta \xi \in \mathcal{N}(T)$ as the 
spatially inhomogeneous dynamics.

In the sequel we analyze separately the exponential
contraction of $\delta \xi$ and the convergence of $\overline{\delta x}$ leading to 
spatially homogeneous motions.
The definition of spatial homogeneity was introduced in 
\cite{aminzare2016}, which we recall in the following lines.
 
\begin{definition}[Spatial homogeneity \cite{aminzare2016}]
  \label{defn:space:uniform}
  The system \eqref{eq:pde:lure}-\eqref{eq:pde:bc} is spatially 
  homogeneous with rate $\mu > 0$ if for any given initial condition
  \begin{equation}
    \label{eq:spatial:uniform}
    \Vert \frac{\partial x}{\partial \theta} (\cdot, t) \Vert_{L_{n}^{2}(\Omega)} \leq M e^{- \mu t} 
    \Vert \frac{\partial x}{\partial \theta} (\cdot, 0) \Vert_{L_{n}^{2}(\Omega)},
  \end{equation}
  where $M > 0$.
\end{definition}

Definition \ref{defn:space:uniform} states that for any initial 
condition the spatial mismatch between any two trajectories, that are infinitesimally close,
decays exponentially to zero, that is, all trajectories converge to each other in the 
spatial domain, enforcing an homogeneous motion in space. Spatial homogeneity is 
thus equivalent to contraction of the spatially inhomogeneous dynamics.

\begin{proposition}
  \label{prop:pde:diff:homogeneity}
  System \eqref{eq:pde:lure}-\eqref{eq:pde:bc} 
  is spatially homogeneous with rate $\mu \geq 0$, if and only if,
  the origin of the system
  \eqref{eq:pde:diff}-\eqref{eq:pde:diff:bc}-\eqref{eq:pde:diff:orthogonal}
  is uniformly exponentially stable with the same rate $\mu \geq 0$.
\end{proposition}

\begin{proof}
  The proof is a direct consequence of Definition \ref{defn:space:uniform}
  and the fact that $\delta \xi = \frac{\partial x}{\partial \theta}$ is the solution to 
  \eqref{eq:pde:diff}-\eqref{eq:pde:diff:bc}-\eqref{eq:pde:diff:orthogonal}.
\end{proof}

	Conditions guaranteeing the exponential homogeneity of 
	\eqref{eq:pde:lure}-\eqref{eq:pde:bc} have been studied extensively, see e.g.,
	 \cite{aminzare2016, arcak2011, conway1978, hale1986}. 
	The dissipativity formulation of those conditions is as follows.
	Let $\sigma: \RE^{m} \times \RE^{m} \to \RE$ be the quadratic differential
	supply rate
	\begin{equation}
		\sigma(\delta y(\theta,t), \delta u(\theta,t)) := \begin{bmatrix}
	    \delta y(\theta, t)
	      \\
	      \delta u(\theta, t)
	    \end{bmatrix}^{\top} 
	    \begin{bmatrix}
	      Q & L
	      \\
	      L^{\top} & R
	    \end{bmatrix}
	    \begin{bmatrix}
	      \delta y(\theta, t)
	      \\
	      \delta u(\theta, t)
	    \end{bmatrix}
	  \label{eq:diff:supply:rate}
	\end{equation}
	where the matrices $Q, L$ and $R$ are constant and $R = R^{\top} \succ 0$.

	\begin{definition}
	  The LTSI system \eqref{eq:pde:sti}-\eqref{eq:pde:bc} 
	  is uniformly differential dissipative with rate $\mu \geq 0$ and with respect to the
	  supply rate \eqref{eq:diff:supply:rate}, if there exists a
	  matrix $\Pi = \Pi^{\top} \succ 0$ such that the following inequality holds for all
	  admissible $\delta u$ with $(\delta x, \delta y)$
	  satisfying \eqref{eq:pde:diff:sti}-\eqref{eq:pde:diff:bc}
	  \begin{multline}
	    \int_{\Omega}
	    \begin{bmatrix}
	      \frac{\partial}{\partial t} \delta x
	      \\
	      \delta x
	    \end{bmatrix}^{\top} 
	    \begin{bmatrix}
	      0 & \Pi 
	      \\
	      \Pi & 2 \mu \Pi + \varepsilon I_{n}
	    \end{bmatrix}
	    \begin{bmatrix}
	      \frac{\partial}{\partial t} \delta x
	      \\
	      \delta x
	    \end{bmatrix} d\theta
	    \leq 
	    \\
	    \int_{\Omega} \sigma(\delta y, \delta u) d \theta
		\label{eq:pde:dissipation}
	  \end{multline}
	  Addtionally, if \eqref{eq:pde:dissipation} holds in an 
	  invariant subspace $\mathcal{V} \subset L_{2}^{n}(\Omega)$ of $\delta x$ then we say
	  that the system is uniformly differential dissipative in $\mathcal{V}$.
	  The property is strict if $\varepsilon > 0$ in \eqref{eq:pde:dissipation}.
	  \label{def:uniform:dissipative}
	\end{definition}

	Henceforth, dissipativity is always asssumed with respect to the supply rate
	\eqref{eq:diff:supply:rate}. With those definitions in place, the dissipativity analysis of 
	spatial homogeneity of \eqref{eq:pde:lure}-\eqref{eq:pde:bc} 
	is an infinite-dimensional version of the classical circle criterion.

	\begin{theorem}
	  Let $\varphi$ satisfy the dissipation inequality 
	  \eqref{eq:dissipation:nonlinearity}.
	  If the LTSI system \eqref{eq:pde:sti}-\eqref{eq:pde:bc} is uniformly differential
	  dissipative with rate $\mu \geq 0$ in $\mathcal{N}(T)$, then the closed-loop
	  system \eqref{eq:pde:lure}-\eqref{eq:pde:bc} is spatially homogeneous with the 
	  same rate $\mu$.
	  \label{thm:dissipation:spatial:homogeneity}
	\end{theorem}

	\begin{proof}
		From Proposition \ref{prop:pde:diff:homogeneity} it follows that spatial
		homogeneity of \eqref{eq:pde:lure}-\eqref{eq:pde:bc} is equivalent to
		exponential stability of the gradient dynamics given by 
		\eqref{eq:pde:diff}-\eqref{eq:pde:diff:orthogonal}. Let $\delta \xi =
		\frac{\partial x}{\partial \theta} \in \mathcal{N}(T)$. By hypothesis, there exists 
		$\Pi = \Pi^{\top} \succ 0$ such that \eqref{eq:pde:dissipation} holds for
		$\delta x = \delta \xi$. Now, let
		$S(\delta \xi) = \langle \delta \xi(\cdot, t), 
		\Pi \delta \xi(\cdot, t) \rangle_{L_{n}^{2}(\Omega)}$, 
		then \eqref{eq:pde:dissipation} is equivalent to
	  \begin{displaymath}
	    \frac{d}{dt} S(\delta \xi) \leq \int_{\Omega} \sigma(\delta y, \delta u) d \theta
	    - 2 \mu S(\delta \xi) - \varepsilon \Vert \delta \xi(\cdot, t) \Vert_{L_{n}^{2}(\Omega)}
	  \end{displaymath}
	Using $\delta u(\theta, t) = - J_{\varphi}(y(\theta, t)) \delta y(\theta, t)$ 
	together with the sector bound \eqref{eq:dissipation:nonlinearity}  
	leads to
	  \begin{equation}
	    \frac{d}{dt} S(\delta \xi) \leq -2 \mu S(\delta \xi) - \varepsilon \Vert 
	    \delta \xi(\cdot, t) \Vert_{L_{n}^{2}(\Omega)}
	    \label{eq:spatial:lyap:ineq:4}
	  \end{equation}
	  Now, multiplying both sides of \eqref{eq:spatial:lyap:ineq:4} by $e^{2 \mu t}$ 
	  and integrating from $\tau = 0$ up to $\tau = t$, yields
	  \begin{displaymath}
	    \Vert \delta \xi(\cdot, t) \Vert_{L_{n}^{2}(\Omega)} \leq 
	    \sqrt{\frac{\lambda_{\max}(\Pi)}{\lambda_{\min}(\Pi)}} e^{-\mu t} 
	    \Vert \delta \xi(\cdot, 0) \Vert_{L_{n}^{2}(\Omega)}
	  \end{displaymath}
	  where $\lambda_{\max}(\Pi)$ and $\lambda_{\min}(\Pi)$ denote 
	  the maximum and minimum eigenvalue of $\Pi$. Therefore, $\delta \xi= \frac{\partial x}{\partial \theta}$ goes
	  exponentially to the zero with rate $\mu \geq 0$.
	\end{proof}

	The following theorem provides a numerical test for
	uniform differential dissipativity of the LTSI system 
	\eqref{eq:pde:sti}-\eqref{eq:pde:bc}.
	The result is a reformulation of 
	\cite[Theorem 1]{arcak2011} for reaction-diffusion systems with periodic 
	spatial domain $S^{1}$.

	\begin{theorem}
	  Let $\Pi \in \RE^{n \times n}$ be a symmetric positive definite matrix such that
	  \begin{equation}
	\label{eq:spatial:uniform:lmi:1}
	\Pi D + D^{\top} \Pi \succeq 0
	  \end{equation}
	  and 
	  \begin{equation}
	\label{eq:spatial:uniform:lmi:2}
	\begin{bmatrix}
	  \Theta_{1,1} & \Pi B - C^{\top} L
	  \\
	  B^{\top} \Pi - L^{\top} C & -R
	\end{bmatrix} \preceq 0
	  \end{equation}
	  where $\Theta_{1,1} = (A - \lambda_{2} D)^{\top} \Pi + \Pi (A - \lambda_{2} D) 
	  + 2 \mu \Pi - C^{\top} Q C + \varepsilon I_{n}$.
	  Then the inhomogeneous dynamics of \eqref{eq:pde:lure}-\eqref{eq:pde:bc} 
	  is uniformly differential dissipative with rate $\mu \geq 0$, 
	  where $\lambda_{q}$ is the $q$-th eigenvalue of the opertor $\Delta$ 
	  with domain \eqref{eq:laplace:domain}, respectively.
	  \label{thm:spatial:uniform:lyapunov}
	\end{theorem}
	\begin{proof}
	  The reader is addressed to \cite{arcak2011} for a detailed proof of this fact.
	\end{proof}

	For the differential spatial dynamics 
	\eqref{eq:pde:diff:orthogonal} implies that $\mathcal{N}(T) = 
	\textrm{span} \{ \nu_{1} \}$, where $\nu_{1}$ is the eigenvector associated 
	to the eigenvalue $\lambda_{1}$.
	Therefore, uniform differential dissipativity of the inhomogenous dynamics
	is tested by solving 
	\eqref{eq:spatial:uniform:lmi:1}-\eqref{eq:spatial:uniform:lmi:2} 
	with $\lambda_{2} = 1$.

	\subsection{Differential homogeneous dynamics}
	The gradient dynamics 
	describes the time evolution of fluctuations in space. The 
	complementary dynamics associated to $\overline{\delta x} := T \delta x$, 
	describes the homogeneous behavior of the differential dynamics.
	Thus, the differential dynamics constrained to $\mathcal{R}(T)$ is constant in space and therefore
	can be described by an ODE. Such intuition is formalized in the following theorem.
	\begin{theorem}
	The dynamics of \eqref{eq:pde:diff}-\eqref{eq:pde:diff:bc} projected into the 
	subspace $\mathcal{R}(T)$ reduces to
	\begin{subequations}
	  \begin{align}
	   & \begin{cases}
	     \frac{d}{dt} \overline{\delta x}(t) = A \overline{\delta x}(t) + B \overline{\delta u}(t) 
	  \\
	  \overline{\delta y}(t) = C \overline{\delta x}(t)
	  \end{cases}
	  \label{eq:ode:lure:diff:linear}
	  \\
	  & \overline{\delta u}(t) = - \left( 
	  \int_{\Omega} J_{\varphi}(y(\theta,t)) d\theta \right) \overline{\delta y}(t)
	    \label{eq:ode:lure:diff:feedback}
	  \end{align}
	    \label{eq:ode:lure:diff}
	\end{subequations}
		\label{thm:emergent:dynamics}
	\end{theorem}

	\begin{proof}
	Applying
	the projection $T$, previously defined in \eqref{eq:projection}, 
	on both sides of \eqref{eq:pde:diff:sti}  yields
	  \begin{equation}
	    \begin{cases}
	     \frac{d}{d t} \overline{\delta x}(t) 
	    = A \overline{\delta x}(t) + B \overline{\delta u}(t)
	    \\
	    \overline{\delta y}(t) = C \overline{\delta x}(t)
	    \end{cases}
		\label{eq:homogeneous:dynamics}
	  \end{equation}
	  where $\overline{\delta x}$ denotes the average in space of the differential variable
	  $\delta x$, introduced in \eqref{eq:projection},
	  similar for $\overline{\delta y}$, and $\overline{\delta u}$.
	  We recall that $\delta x = \overline{\delta x} + \delta \xi$, where $\delta \xi = 
	  \frac{\partial x}{\partial \theta} \in \mathcal{N}(T)$.
	  Hence, $\overline{\delta u}(t)$ obeys,
	  \begin{align*}
		  \overline{\delta u}(t)
		  & = 
		  -\int_{\Omega} \left( J_{\varphi}(y(\theta,t)) \delta y(\theta,t) 
		  \right) d \theta
		  \\
		  & =
		  -\int_{\Omega} \left( \frac{\partial}{\partial \theta} \varphi(y(\theta,t))
		  + J_{\varphi}(y(\theta,t)) \overline{\delta y}(t) \right) d\theta
		  \\
		  & =
		  - \left( \int_{\Omega} J_{\varphi}(y(\theta,t))  d \theta
		  \right) \overline{\delta y}(t)
	  \end{align*}
	  where we used the fact that $\delta y = C \overline{\delta x} + C \delta \xi$ in the 
	  second equation, and the fundamental theorem of calculus together with
	  \eqref{eq:pde:bc} to obtain the last equation.
	  \end{proof}

	The above result agrees with the traditional approach of \cite{conway1978, hale1986} 
	in which spatial homogeneity reduces a PDE into an ODE.
	Thus, \eqref{eq:ode:lure:diff} describes the differential dynamics of the 
	homogeneous behavior which we identify as the differential temporal dynamics.

	We now illustrate the use of differential dissipativity analysis to study 
	non-equilibrium asymptotic  behaviors of the homogeneous dynamics. 
	We make use of a recent development of the theory in 
	\cite{forni2019, Miranda2017b, miranda2020}.
	We recall that the inertia of a symmetric matrix $P \in \RE^{n \times n}$ 
	is the triple $(\nu, \zeta, \pi)$ where each entry denotes the number of 
	negative, zero and positive eigenvalues, respectively. 
	
	Note that for the homogeneous dynamics, the associated supply rate	
	is the same as \eqref{eq:diff:supply:rate} but with functions that
	are independent of the spatial variable. Namely,  it follows from 
	\eqref{eq:dissipation:nonlinearity} that
	
	\begin{align}
	  \nonumber
	  0 & \geq
	  \langle \overline{\delta y}, Q \overline{\delta y} \rangle_{L_{n}^{2}(\Omega)}
	+  \langle ( J_{\varphi}
	  \circ y) \overline{\delta y}, R \left( J_{\varphi} \circ y \right) \overline{\delta y} 
  	  \rangle_{L_{n}^{2}(\Omega)}
	  \\
	\nonumber
	  & -\langle \overline{\delta y}, L (J_{\varphi} \circ y ) \overline{\delta y} 
	  \rangle_{L_{n}^{2}(\Omega)}
	  - \langle (J_{\varphi} \circ y) \overline{\delta y},
	  L^{\top} \overline{\delta y} \rangle_{L_{n}^{2}(\Omega)}
	  \\
	  \label{eq:lumped:supply}
	  & \geq
	  \begin{bmatrix}
		 \overline{\delta y}(t) \\ \overline{\delta u}(t)
	  \end{bmatrix}^{\top}
	  \begin{bmatrix}
		  Q & L
		  \\
		  L^{\top} & R
	  \end{bmatrix}
	  \begin{bmatrix}
		  \overline{\delta y}(t) \\ \overline{\delta u}(t)
	  \end{bmatrix} =: \bar{\sigma}(\overline{\delta y}, \overline{\delta u})
	  \end{align}
	  where we applied Jensen's inequality to the quadratic term 
	  $\langle \eta, R \eta \rangle_{L_{n}^{2}(\Omega)}$ to obtain the last inequality.

	  Associated to the variational dynamics \eqref{eq:ode:lure:diff} we consider
	  the family of lumped systems
	  \begin{subequations}
	  \begin{align}
	   & \begin{cases}
	     \frac{d}{dt} \bar{x}(t) = A \bar{x}(t) + B \bar{u}(t)
	  \\
	  \bar{y}(t) = C \bar{x}(t)
	  \end{cases}
	  \label{eq:ode:lure:linear}
	  \\
	  & 
	  \begin{bmatrix}
		  \overline{\delta y}(t) \\ \overline{\delta u}(t)
	  \end{bmatrix}^{\top}
	  \begin{bmatrix}
		  Q & L
		  \\
		  L^{\top} & R
	  \end{bmatrix}
	  \begin{bmatrix}
		  \overline{\delta y}(t) \\ \overline{\delta u}(t)
	  \end{bmatrix} \leq 0
	    \label{eq:ode:lure:feedback}
	  \end{align}
	    \label{eq:ode:lure}
	\end{subequations}
	where $\bar{u}$ is defined implicitly through its variational representation 
	$\overline{\delta u}$. 

	\begin{definition}
	  The linear system \eqref{eq:ode:lure:linear} is $p$-dissipative with rate
	  $\lambda \geq 0$ and with respect to the supply rate $\bar{\sigma}$,
	  if there exists a symmetric matrix $P = P^{\top}$ with
	  inertia $(p, 0, n-p)$ such that for all admissible $\overline{\delta u}$ and all
	  $(\overline{\delta x}, \overline{\delta y})$ 
	  satisfying \eqref{eq:ode:lure:diff} the following holds
	  \begin{equation}
	    \begin{bmatrix}
	      \frac{d}{dt} \overline{\delta x} \\ \overline{\delta x}
	    \end{bmatrix}^{\top}
	    \begin{bmatrix}
	      0 & P
	      \\
	      P & 2 \lambda P + \varepsilon I_{n}
	    \end{bmatrix}
	    \begin{bmatrix}
	      \frac{d}{dt} \overline{\delta x} \\ \overline{\delta x}
      \end{bmatrix} \leq \bar{\sigma}(\overline{\delta y}, \overline{\delta u})
	  \label{eq:p:dissipative}
	  \end{equation}
	  \label{defn:p:dissipative}
	  The property is strict if $\varepsilon > 0$.
	\end{definition}

	Analogous to the inhomogeneous case above, in what follows, we consider only 
	quadratic supply rates of the form \eqref{eq:lumped:supply}.
	The following theorem, taken from \cite{Miranda2017b} and repeated here for 
	completeness,
	provides useful information for characterizing the homogenous part of the
	asymptotic behavior. 

	\begin{theorem}
	  Let $\varphi$ satisfy the dissipation inequality 
	  \eqref{eq:dissipation:nonlinearity}.
	  If the system \eqref{eq:ode:lure:linear} is strictly $p$-dissipative with 
	  rate $\lambda \geq 0$. 
	  Then the homogeneous dynamics of the closed-loop
	  \eqref{eq:pde:lure}-\eqref{eq:pde:bc} is $p$-dominant. 
	  In particular, each bounded solution
	 asymptotically  converges to an equilibrium for $p=1$ and to a simple limit set 
	 (equilibrium, closed orbit, or connected arc of equilibria) for $p=2$.
	  \label{thm:dominance:ode}
	\end{theorem}
	\begin{proof}
	  The homogeneous differential dynamics of the closed-loop 
	  \eqref{eq:pde:lure}-\eqref{eq:pde:bc}
	  is given by \eqref{eq:ode:lure:diff}, which is a lumped Lur'e system, 
	  Theorem \ref{thm:emergent:dynamics}. The result thus follows from 
	  \cite[Theorem 4.2]{Miranda2017b}. 
	\end{proof}

	It follows from Definition \ref{defn:p:dissipative} that 
	the homogeneous dynamics associated to \eqref{eq:pde:lure}-\eqref{eq:pde:bc}
	is strictly $p$-dissipative with 
	rate $\lambda \geq 0$ if there exist $\varepsilon > 0$ and a matrix 
	$P = P^{\top}$ with inertia $(p, 0, n-p)$ satisfying
	  \begin{equation}
	    \begin{bmatrix}
	      \hat{\Theta}_{1,1} & PB - C^{\top} L
	    \\
	    B^{\top} P - L^{\top} C & -R
	    \end{bmatrix} \preceq 0
		\label{eq:lmi:p:dissipative}
	  \end{equation}
	  where $\hat{\Theta}_{1,1} = A^{\top} P + P A + 2 \lambda P - C^{\top} Q C 
	  + \varepsilon I_{n}$

	In this way, the differential model \eqref{eq:pde:diff}-\eqref{eq:pde:diff:bc}
	contains all the information needed for the study of the global behavior of 
	\eqref{eq:pde:lure}-\eqref{eq:pde:bc}.

	\begin{example}
	  We illustrate the above analysis with an application to the 
	  Nagumo model describing the 
	  spatio-temporal dynamics of a bistable transmission line \cite{nagumo1965},
	  \begin{equation}
	    \frac{\partial x}{\partial t}(\theta,t) = D \Delta x(\theta,t) + A x(\theta, t) - \varphi(x(\theta,t))
	    \label{eq:pde:nagumo}
	  \end{equation}
	  where $x(\theta,t) \in \RE$, $D > 0$, and $\varphi: \RE \to \RE$ is an ``$N$-shape'' 
	  function as the one shown in Figure \ref{fig:n:shape}. Thus, $\varphi$ satisfies 
	  \eqref{eq:diff:sector} for some $K_{1} < 0 < K_{2}$.
	  The boundary conditions are the same as in \eqref{eq:pde:bc}.
	  
	  \begin{figure}[htpb]
	    \centering
	    \includegraphics[trim={0 5mm 0 1mm}, clip]{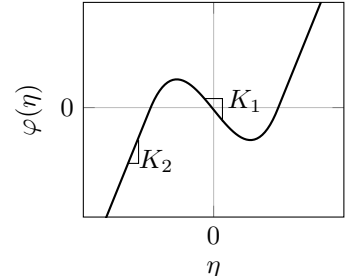}
	    \caption{``$N$-shape'' nonlinear function in the differential sector 
	    $[K_{1}, K_{2}]$.}
	    \label{fig:n:shape}
	  \end{figure}
	  In this example, condition \eqref{eq:spatial:uniform:lmi:1} reduces to $\Pi > 0$ and 
	  by using Schur's complement formula it follows 
	  that \eqref{eq:spatial:uniform:lmi:2} is equivalent to the  condition,
	  \begin{displaymath}
	    \Pi^{2} + 2 \left(A + \mu - D - \frac{K_{1}+K_{2}}{2} \right) \Pi  
	    + \frac{(K_{1} - K_{2})^{2}}{4} < 0
	  \end{displaymath}
	  Straightforward computations reveal that uniform differential dissipativity 
	  of the inhomogeneous dynamics with rate at least $\mu$ is guaranteed whenever
	  \begin{equation}
	    D > A + \mu - K_{1}
	    \label{eq:nagumo:examp:diffusion}
	  \end{equation}
	  which implies spatial homogeneity of the closed-loop 
	  \eqref{eq:pde:lure}-\eqref{eq:pde:bc} according to 
	  Theorem \ref{thm:dissipation:spatial:homogeneity}.
	  Now, the complementary dynamics in $\mathcal{R}(T)$ is given by 
	  \eqref{eq:ode:lure:diff}, whose dissipativity property is verified by 
	  \eqref{eq:lmi:p:dissipative}, which in this case reduces into
	  \begin{equation}
	    P^{2} + 2 \left( A + \lambda - \frac{K_{1} +
	    K_{2}}{2} \right) P + \frac{(K_{1} - K_{2})^{2}}{4} < 0
	    \label{eq:ex1:temporal:lmi}
	  \end{equation}
	  It is easy to verify that if $K_{1} > A$ then \eqref{eq:ex1:temporal:lmi} admits a 
	  positive solution $P > 0$, that is, the  homogeneous dynamics 
	  is $0$-dissipative with rate $0 < \lambda < K_{1} - A$. 
	  In such case, there is a unique equilibrium for
	  \eqref{eq:pde:lure}-\eqref{eq:pde:bc} that is globally asymptotically stable,
	  that is, the complete spatio-temporal behavior goes towards the unique equilibrium.
	  On the other hand, if $A > K_{1}$, then \eqref{eq:ex1:temporal:lmi} admits a 
	  negative solution $P < 0$, that is, the homogeneous dynamics 
	  is $1$-dissipative with positive rates $\lambda$ 
	  satisfying $\lambda > K_{2} - A$.
	  Further, from a conventional local stability analysis one gets that the origin 
	  of the dynamics in $\mathcal{R}(T)$ is unstable whenever $A > K_{1}$. 
	  Thus, when condition \eqref{eq:nagumo:examp:diffusion} and $A > K_{1}$ hold, then
	  the PDE \eqref{eq:pde:nagumo} will have a homogeneous bistable behavior. 
	  Figure \ref{fig:nagumo:example} shows the 
	  spatio-temporal evolution of the system to two different initial conditions with the following parameters $A = 0$, $D = 1.1$, 
	  $K_{1} = -1$, and $K_{2} = 1$.

	  \begin{figure}[htpb]
	    \centering
	    \includegraphics[width = 0.9\columnwidth, trim={6mm 3mm 5mm 1mm}, clip]
	    {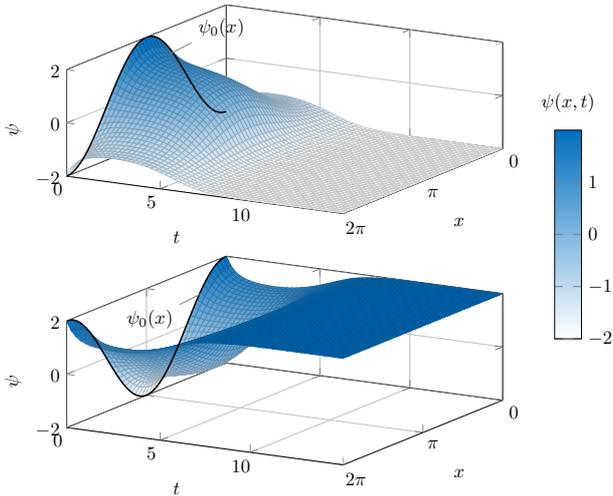}
	    \caption{Spatio-temporal evolution of trajectories of Nagumo's equation \eqref{eq:pde:nagumo} 
	    to two different initial conditions showing both, the spatial homogeneity of 
	    solutions and the bistable nature of the transmission line.}
	    \label{fig:nagumo:example}
	  \end{figure}
	\end{example}

	\section{Analysis in the frequency domain}
	\label{sec:freq}

	The linear system  \eqref{eq:pde:lure}-\eqref{eq:pde:bc} is both 
	space and time invariant (LTSI): solutions shifted in time and in space
	satisfy the same equation \cite{bamieh2002}.

	Spatial and temporal invariance properties of linear systems allow for insightful 
	frequency domain analysis. In this section, we briefly illustrate the frequency-domain 
	interpretation of the results of the previous sections. 

	\subsection{Differential inhomogeneous dynamics}

	Spatial invariance allows to analyze a linear
	PDE as a family of ODEs parametrized by the spatial frequency $\zeta$ 
	\cite{ayres2002, bamieh2002, curtain2009, gorinevsky2008}. 
	By taking the Fourier transform of \eqref{eq:pde:diff:sti} with 
	respect to the spatial variable $\theta$, we transform the PDE 
	\eqref{eq:pde:diff:sti} into the family of linear systems
	\begin{equation}
	  \label{eq:ode:family}
	  \begin{cases}
	    \frac{d}{dt} \delta x_{\zeta}(t) = 
	    (A - \zeta^{2} D) \delta x_{\zeta}(t) + B \delta u_{\zeta}(t)
	  \\
	  \delta y_{\zeta}(t) = C \delta x_{\zeta}(t)
	  \end{cases}
	  \end{equation}
	  where, for the case of $\Omega = S^{1}$, 
	  $\zeta \in \mathbb{Z}$ (the dual group to $S^{1}$).
	Notice that each $\delta x_{\zeta}(t)$, $\zeta \in \mathbb{Z}$, is a
	coefficient on the Fourier series expansion of $\delta x(\cdot, t)$, \cite{curtain2009}.

	The splitting between spatial and temporal differential dynamics in the previous
	section has an obvious interpretation in the frequency domain:  (\ref{eq:ode:family})
	reduces to the differential temporal dynamics for the uniform spatial mode, 
	that is $\zeta = 0$, whereas the differential spatial dynamics correspond to
	all other modes $\zeta \in \mathbb{Z} \setminus \{0\}$.

	The following theorem provides sufficient conditions that guarantee the 
	spatial homogeneity of the closed-loop 
	\eqref{eq:pde:lure}-\eqref{eq:pde:bc} via the 
	family of ODEs \eqref{eq:ode:family}.

	\begin{theorem}
	Suppose that for each $\zeta \in \mathbb{Z} \setminus \{0\}$, the linear system
	  \eqref{eq:ode:family} is $0$-dissipative with rate $\mu \geq 0$ and with the
	  same storage function $S(\delta x_{\zeta}) = \delta x_{\zeta}^{\top} \Pi
	  \delta x_{\zeta}$. Then the closed-loop
	  \eqref{eq:pde:lure}-\eqref{eq:pde:bc} is spatially homogeneous with the same rate 
	  $\mu$.
	  \label{thm:spatial:freq:dissipative:uniform}
	\end{theorem}

	\begin{proof}
	  The hypothesis on the family of systems \eqref{eq:ode:family} is equivalent to the
	  existence of a matrix $\Pi = \Pi^{\top} \succ 0$ satisfying
	  the following family of parametrized LMIs
	  \begin{equation}
	    \Theta_{\zeta}(\Pi) := 
	    \begin{bmatrix}
	      \Theta_{1,1}(\zeta) & \Pi B - C^{\top} L
	    \\
	    B^{\top} \Pi  - L^{\top} C & -R
	    \end{bmatrix} \preceq 0
		\label{eq:ode:family:lmi}
	  \end{equation}
	were $ \Theta_{1,1}(\zeta) = (A - \zeta^{2} D)^{\top} \Pi  + \Pi  (A - \zeta^{2} D) + 
	2 \mu \Pi - C^{\top} Q C + \varepsilon I_{n} $.
	The rest of the proof consists in showing that \eqref{eq:ode:family:lmi} 
	is equivalent to conditions 
	\eqref{eq:spatial:uniform:lmi:1}-\eqref{eq:spatial:uniform:lmi:2}.
	To that end, let $\tau = \frac{1}{\zeta^{2}} \in (0, 1]$. It then follows that 
	condition \eqref{eq:ode:family:lmi} holds for all
	  $\zeta \in \mathbb{Z} \setminus\{0\}$ if and only if 
	  \begin{equation}
	\label{eq:ode:family:lmi:convex}
	    \begin{bmatrix}
	      \tilde{\Theta}_{1,1}(\tau) & \tau \left( \Pi B - C^{\top} L  \right)
	      \\
	      \tau \left( B^{\top} \Pi - L^{\top} C \right) &  -\tau R
	    \end{bmatrix} \preceq 0
	  \end{equation}
	  holds for all $\tau \in (0, 1]$, where $\tilde{\Theta}_{1,1,}(\tau) = (\tau A- D)^{\top} \Pi + \Pi ( \tau A - D )
	  - \tau (C^{\top} Q C + 2 \mu \Pi + \varepsilon I_{n})$. Now, let us assume first that \eqref{eq:ode:family:lmi:convex}
	  holds. Thus, setting $\tau = 1$ in  
	  \eqref{eq:ode:family:lmi:convex} implies \eqref{eq:spatial:uniform:lmi:2}. Next, a necessary condition
	  for \eqref{eq:ode:family:lmi:convex} to hold is 
	  \begin{displaymath}
	    -D^{\top} \Pi - \Pi D + \tau (A^{\top} \Pi + \Pi A - C^{\top} Q C + 2 \mu \Pi + \varepsilon I_{n}) 
	     \preceq 0
	  \end{displaymath}
	  for all $\tau \in (0, 1]$. Such condition is possible only if \eqref{eq:spatial:uniform:lmi:1} holds.
	  The converse statement follows directly by noting that \eqref{eq:ode:family:lmi:convex} is contained in the
	  convex combination of conditions 
	  \eqref{eq:spatial:uniform:lmi:1}-\eqref{eq:spatial:uniform:lmi:2}. 
	  Hence \eqref{eq:ode:family:lmi} implies uniform differential dissipativity
	  of \eqref{eq:pde:sti}-\eqref{eq:pde:bc} with rate $\mu$ on $\mathcal{N}(T)$ 
	  and the conclusion follows from Theorem \ref{thm:dissipation:spatial:homogeneity}.
	\end{proof}

	\begin{remark}
	\label{remark:frequency}
	The LMI \eqref{eq:ode:family:lmi}  has the interpretation of a dissipativity analysis
	of the family of systems \eqref{eq:ode:family} 
	in feedback interconnection with a family of $\zeta$-parametrized time-varying gains 
	$\delta u_{\zeta} =-\tilde{J}_{\zeta}(t) \tilde{y}_{\zeta}$ satisfying
	\begin{equation}
	    \begin{bmatrix}
	      I_{m} \\ -\tilde{J}_{\zeta}(t)
	   \end{bmatrix}^{\top}
	   \begin{bmatrix}
	     Q & L
	     \\
	     L^{\top} & R
	   \end{bmatrix}
	   \begin{bmatrix}
	     I_{m} \\ -\tilde{J}_{\zeta} (t)
	   \end{bmatrix}
	   \preceq 0.
	   \label{eq:nonlinear:freq}
	\end{equation}
	For each value of $\zeta$, the storage $S(\delta x_{\zeta}) = 
	\delta x_{\zeta}^{\top} \Pi \delta x_{\zeta}$, where $\Pi = \Pi \succeq 0$
	satisfies
	\begin{multline*}
	  \frac{d}{dt} S(\delta x_{\zeta}) = 
	  \\
	  \begin{bmatrix}
	    \delta x_{\zeta} \\ \delta u_{\zeta}
	  \end{bmatrix}^{\top}
	  \begin{bmatrix}
	    (A - \zeta^{2} D)^{\top} \Pi + \Pi (A - \zeta^{2} D) & \Pi B
	    \\
	    B^{\top} \Pi & 0
	  \end{bmatrix}
	  \begin{bmatrix}
	    \delta x_{\zeta} \\ \delta u_{\zeta}
	  \end{bmatrix}
	\end{multline*}
	  and the application of the $S$-procedure yields \eqref{eq:ode:family:lmi} as a sufficient condition for the uniform
	  exponential stability of the family of closed-loops. 
	  It is worth stressing that in general $\tilde{J}_{\zeta}(t) \delta y_{\zeta}(t)$ is {\it not} the spatial Fourier transform of
	  the term $J_{\varphi}(y(\theta,t)) \delta y(\theta,t)$ in \eqref{eq:pde:diff:feedback}. 
	\end{remark}

	In the previous subsection we analyzed spatial homogenity via the LMIs 
	\eqref{eq:spatial:uniform:lmi:1}-\eqref{eq:spatial:uniform:lmi:2}. 
	The analysis in the spatial frequency domain in this section provides
	an alternative: because the Fourier 
	transform is an isometry between $L_{n}^{2}(\Omega)$ and $l_{n}^{2}(\mathbb{Z})$, 
	it is sufficient to show that the dynamics of each Fourier coefficient, 
	given by \eqref{eq:ode:family}, converges exponentially to zero with rate 
	at least $\mu$ for each $\zeta \in \mathbb{Z} \setminus \{0\}$. 
	Therefore, it is enough to verify the stability of \eqref{eq:ode:family} 
	subject to the quadratic constraint $\sigma(\delta y_{\zeta}, \delta u_{\zeta}) \leq 0$.
	That is, to verify only the individual dissipativity properties of each 
	Fourier coefficient.
	To this end, let us introduce the family of
	transfer functions associated to 
	\eqref{eq:ode:family} as 
	\begin{equation}
	  G_{\zeta}(s) = C \left( sI - (A - \zeta^{2}D) \right)^{-1} B
	  \label{eq:transfer:function}
	\end{equation}
	where $s \in \mathbb{C}$ and $\zeta \in \mathbb{Z} \setminus \{0\}$.
	In the SISO case, graphical tests (circle criterion) can be derived. 
	Let $\mathcal{D}(K_{1}, K_{2})$ be the disk in the complex plane given by the set
	\begin{multline}
	  \mathcal{D}( K_{1}, K_{2}) := \left\{ x + jy \in \mathbb{C} \big| \left( x + \frac{K_{1} + K_{2}}{2 K_{1} K_{2}}^{2} \right)
	    + y^{2} \right.
	   \\
	  \leq \left. \left( \frac{K_{2} - K_{1}}{2 K_{1} K_{2}} \right)^{2} \right\}
	  \label{eq:disk}
	\end{multline}

	\begin{theorem}
	  \label{thm:spatial:circle}
	  Let $\varphi: \RE^{m} \to \RE^{m}$ be such that it satisfies the differential sector condition \eqref{eq:diff:sector}.
	  If for each $\zeta \in \mathbb{Z} \setminus \{0\}$ there exists $\mu \geq 0$ such that
	  \begin{enumerate}
	    \item $G_{\zeta}(s - \mu)$ has no poles on the closure of $\mathbb{C}_{+}$;
	    \item  one of the following conditions is satisfied
	      \begin{enumerate}
		\item $0 < K_{1} < K_{2}$ and the Nyquist plot of $G(s - \mu)$ lies outside the disk 
		  $\mathcal{D}(K_{1}, K_{2})$.
		\item $K_{1} < 0 < K_{2}$ and the Nyquist plot of $G(s - \mu)$ lies inside the disk 
		  $\mathcal{D}(K_{1}, K_{2})$.
		\item $K_{1} < K_{2} < 0$ and the Nyquist plot of $G(s - \mu)$ lies outside the disk 
		  $\mathcal{D}(K_{1}, K_{2})$.
	      \end{enumerate}
	  \end{enumerate}
	  Then the closed-loop \eqref{eq:pde:lure}-\eqref{eq:pde:bc} is spatially homogeneous
	  with rate $\mu$.
	\end{theorem}

	\begin{proof}
	  The proof is the same as in the standard circle criterion, see e.g., 
	  \cite{haddad2008, khalil2002}.
	\end{proof}

	\begin{remark}
	  It is worth to stress that in Theorem \eqref{thm:spatial:circle} we have disregarded
	  the cases in which the Nyquist plot make encirclements of the disk 
	  $\mathcal{D}(K_{1}, K_{2})$. This is because the Riemann-Lebesgue lemma 
	  \cite[p. 36]{conway1990} states that $\delta x_{\zeta} 
	  \to 0$ as $\vert \zeta \vert \to + \infty$. Therefore,
	  the family of Nyquist plots cannot make encirclements of any given disk.
	  \label{remark:circle}
	\end{remark}

	\subsection{Differential homogeneous dynamics}

	The second part of the analysis concerns the asymptotic behavior
	of the model \eqref{eq:ode:lure:diff}, which is finite dimensional.
	In such case the frequency domain approach is explored 
	in \cite{Miranda2017b}, where sufficient conditions are guaranteed.

	The analysis is now centered around the feedback interconnection
	of \eqref{eq:ode:family} with $\zeta = 0$ and a nonlinear term 
	$\varphi: \RE^{m} \to \RE^{m}$
	satisfying \eqref{eq:diff:sector}.
	For the sake of completeness we state the main result for the case of SISO systems,
	whose proof can be found in \cite{Miranda2017b}.

	\begin{theorem}
	  [Extended circle criterion]
	  Consider the closed-loop system \eqref{eq:ode:lure}. 
	  Let $G_{0}(s)$ be the transfer function associated to \eqref{eq:ode:lure:linear} and let $\varphi: \RE \to \RE$ 
	  satisfy the differential sector condition \eqref{eq:diff:sector}. 
	  Then the homogeneous dynamics is
	  $p$-dominant with rate $\lambda \geq 0$ if
	  \begin{enumerate}
	    \item $G_{0}(s - \lambda)$ has $q$ poles on the interior of $\mathbb{C}_{+}$ and no poles on the $j \omega$-axis;
	    \item The Nyquist plot of $G_{0}(s - \lambda)$ makes $E = p - q$ clockwise encirclements of the 
	  point $-1/K_{1}$;
	\item one of the following conditions is satisfied
	  \begin{enumerate}
		\item $0 < K_{1} < K_{2}$ and the Nyquist plot of $G(s - \lambda)$ lies outside the disk 
		  $\mathcal{D}(K_{1}, K_{2})$.
		\item $K_{1} < 0 < K_{2}$ and the Nyquist plot of $G(s - \lambda)$ lies inside the disk 
		  $\mathcal{D}(K_{1}, K_{2})$.
		\item $K_{1} < K_{2} < 0$ and the Nyquist plot of $G(s - \lambda)$ lies outside the disk 
		  $\mathcal{D}(K_{1}, K_{2})$.
	      \end{enumerate}
	\end{enumerate}
	  \label{cor:circle:p:dominant}
	\end{theorem}

	Theorem 
	\ref{thm:spatial:circle} gives us a
	sufficient condition for spatial homogeneity of  
	reaction-diffusion systems, whereas Theorem \ref{cor:circle:p:dominant}
	gives us a sufficient condition for the type of homogeneous motion.

	\begin{example}
	  We apply our approach to the FitzHugh-Nagumo equation
	  \begin{equation}
	    \begin{cases}
	      \frac{\partial x_{1}}{\partial t}(\theta, t) = D_{1,1} \Delta x_{1}(\theta,t) - x_{2}(\theta,t) + u(\theta,t)    \\
	    \varepsilon \frac{\partial x_{2}}{\partial t}(\theta,t) = D_{2,2} \Delta x_{2}(\theta,t) + a x_{1}(\theta,t) - b x_{2}(\theta,t)
	    \\
	    y(\theta,t) = x_{1}(\theta,t)
	    \\
	    u(\theta,t) = - \varphi(y(\theta,t))
	    \end{cases}
	    \label{eq:ex2:fn}
	      \end{equation}
	      where $\varphi: \RE \to \RE$ is a nonlinear ``$N$-shape'' 
	      function in the differential sector $[K_{1}, K_{2}]$,
	      as the one shown in Figure \ref{fig:n:shape}. We first focus on the 
	      analysis of spatial homogeneity.
	     The family of transfer functions
	      $G_{\zeta}(s)$ has the form
	      \begin{equation}
		G_{\zeta}(s) = 
		\frac{s + \frac{1}{\varepsilon} (b + \zeta^{2} D_{2,2,} )}{\left( s 
		  + \zeta^{2} D_{1,1} \right)
		\left( s + \frac{1}{\varepsilon}(b + \zeta^{2} D_{2,2}) \right) + 
	      \frac{a}{\varepsilon}}
	      \label{eq:ex2:tf:spatial}
	      \end{equation}
	      
	      Now, we check the conditions stated in Theorem \ref{thm:spatial:circle}.
	      Thus, if 
	      \begin{displaymath}
		\mu < \min \left\{ D_{1,1}, \frac{1}{\varepsilon}(b + D_{2,2}) \right\}
	      \end{displaymath}
	      then condition 1 holds for all $\zeta \in \mathbb{Z} \setminus \{0\}$. 
	      
	      Setting the parameters as, $D_{1,1} = 0.5$, $D_{2,2} = 0.02$, 
	      $\varepsilon = 0.1$, $a = 0.1$, $b = 0.05$, 
	      $K_{1} = -1.0$ and $K_{2} = 1.0$, we now look 
	      for the values of $\mu$ for which condition 2-(b) also
	      holds. Thus, setting $\mu = 0.01$, we get the family of
	      Nyquist plots depicted in Figure \ref{fig:spatial:nyquist:FN}.
	      \begin{figure}[!htpb]
		\centering
		\includegraphics[width = 0.7\columnwidth, trim={3mm 3mm 3mm 3mm}, clip]{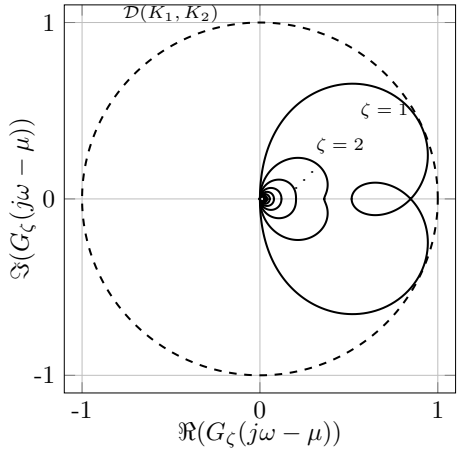}
		\caption{Family of Nyquist plots of \eqref{eq:ex2:tf:spatial} for 
		  $\zeta \in \mathbb{Z} \setminus \{0\}$ and with parameters $D_{1,1} = 0.5$, 
		  $D_{2,2} = 0.02$, $\varepsilon = 0.1$, $a = 0.1$, $b=0.05$ and $\mu = 0.01$.}
		\label{fig:spatial:nyquist:FN}
	      \end{figure}

	      From Figure \ref{fig:spatial:nyquist:FN} it follows that, 
	      with our choice of parameters, we can expect a rate of convergence 
	      of the synchronization error of at least $\mu = 0.01$.
	      With that information on the rate $\mu$, we verify a solution to the 
	      LMI conditions \eqref{eq:spatial:uniform:lmi:1}-\eqref{eq:spatial:uniform:lmi:2} 
	      and we get a positive definite solution $\Pi$ as
	      \begin{displaymath}
		\Pi = 
		\begin{bmatrix}
		  1.16451 & -0.61023
		  \\
		  -0.61023 & 1.1594
		\end{bmatrix}
	      \end{displaymath}
		which confirms the spatial 
		homogeneity of the FitzHugh-Nagumo equation with the selected parameters.
	      
		The following step consists in retrieving the type of synchronized motion. 
		To that end, we make use of 
		the extension of the circle criterion on Theorem \ref{cor:circle:p:dominant}.
		The transfer function of interest is
		\begin{displaymath}
		  G_{0}(s) = \frac{s + \frac{b}{\varepsilon}}{s^{2} + \frac{b}{\varepsilon} s + \frac{a}{\varepsilon}}
		\end{displaymath}
		whose poles are at $\left\{ \frac{-b \pm \sqrt{b^{2} - 
		4 a \varepsilon}}{2 \varepsilon} \right\}$.
		Now, we proceed to verify assumptions 1-3 in 
		Theorem \ref{cor:circle:p:dominant}. First, 
		with our choice of parameters, we have that $b^{2} - 4 a \varepsilon = -0.0375 < 0$.  
		It follows that for any $\lambda > \frac{b}{2 \varepsilon} = 0.25 $, assumption 1 is 
		satisfied with $q = 2$. Selecting $\lambda = 0.8$, we get the Nyquist diagram of Figure \ref{fig:temporal:nyquist:FN}
		from which, we verify condition 2, (with $E = 0$ and therefore $p = 2$), and 3-(b). Hence, the homogeneous dynamcis
		is $2$-dominant.
		\begin{figure}[!t]
		\centering
		\includegraphics[width = 0.7\columnwidth, trim={3mm 3mm 3mm 3mm}, clip]{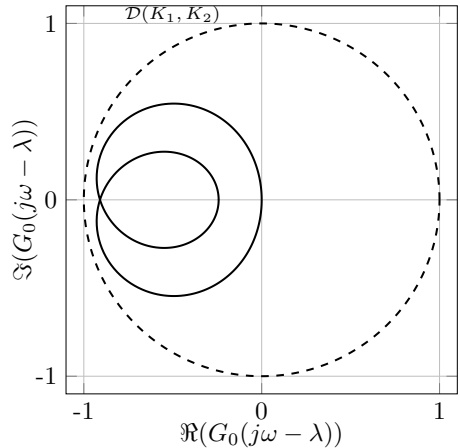}
		\caption{Nyquist plot of the transfer function $G_{0}(s - \lambda)$ associated to system \eqref{eq:ode:lure:diff:linear} 
		with temporal rate $\lambda = 0.8$.}
		\label{fig:temporal:nyquist:FN}
	      \end{figure}

	      Further analysis shows that the origin of the closed-loop \eqref{eq:ode:lure} is the unique equilibrium point and it is unstable.
	      Hence, a cohesive oscillatory behavior is expected. Figure \ref{fig:fitzhugh:nagumo:osc} confirms the analysis.
	      \begin{figure}[!t]
		\centering
		\includegraphics[width = 0.9\columnwidth, trim={6mm 3mm 5mm 0}, clip]{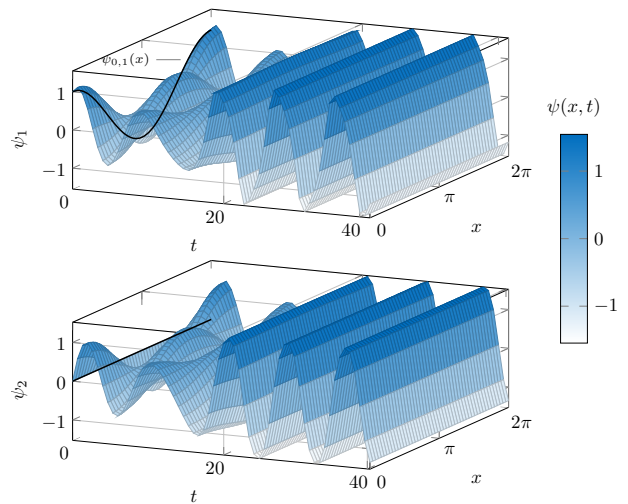}
		\caption{Spatio-temporal evolution of state trajectories 
		of FitzHugh-Nagumo model \eqref{eq:ex2:fn} showing an homogeneous oscillatory
	      behavior.}
		\label{fig:fitzhugh:nagumo:osc}
	      \end{figure}
	\end{example}

It is noteworthy that the methods presented here also find application in the analysis
of systems with compact spatial domain and Neumann boundary conditions, 
as stated in \cite{armbruster1987}. 
Roughly speaking, let $F(x) = D \Delta x + A x - B \varphi(C x)$, then
$F$ is equivariant if 
\begin{displaymath}
	T_{\vartheta} \circ F = F \circ T_{\vartheta}, \text{ and }
	R \circ F =  F \circ R
\end{displaymath} 
where $T_{\vartheta} x(\theta,t) = x(\theta + \vartheta, t)$ and $R x(\theta,t) = 
x(-\theta,t)$. Thus, for equivariant equations, any solution satisfying 
Neumann boundary conditions can be extended, by reflection around the origin, 
to a solution of the periodic problem. 
Additionally, the even part of solutions to the periodic problem are also 
solutions to the Neumann problem, see \cite{armbruster1987} for details. 
Thereby, our approach also extends to the analysis of 
reaction-diffusion systems with Neumann constraints.

	\section{Conclusions}
	\label{sec:end}

	We  illustrated the potential of differential dissipativity for the analysis
	of nonlinear reaction-diffusion systems. The differential dynamics naturally 
	decompose  into two components, the differential 
	inhomogeneous dynamics and the differential homogeneous dynamics. 
	We illustrated sufficient conditions
	for spatial homogeneity, that is, contraction of the differential 
	inhomogeneous dynamics, and for $p$-differential
	dissipativity of the differential homogeneous dynamics. Future work will explore the same framework
	to analyze asymptotic spatiotemporal behaviors that are homogeneous neither in space nor in time.
	Such behaviors include traveling waves and spatiotemporal patterns.

\bibliography{Biblio.bib} 
\bibliographystyle{plain}

\end{document}